\documentclass[a4paper,10pt]{article}
\usepackage{amsfonts}
\usepackage{amssymb}
\usepackage{latexsym}
\usepackage{enumerate}
\usepackage{amsmath}
\usepackage{amsthm}
\usepackage{graphicx}
\usepackage{multirow}
\usepackage{bbm}
\usepackage[export]{adjustbox}
\usepackage{color}

\begin{document}

\title{Systemic risk assessment \\ through high order clustering coefficient}
\author{Roy Cerqueti$^{\flat}$, Gian Paolo Clemente$^{\natural}$,  Rosanna Grassi$^{\sharp}\thanks{Corresponding author.}$\\
 $^{\flat}$ {\small University of Macerata, Department of Economics and Law. }\\{\small Via Crescimbeni 20, 62100, Macerata,
Italy.}\\ {\small Tel.: +39 0733 2583246; fax: +39 0733 2583205. Email: roy.cerqueti@unimc.it} \\
$^{\natural}$  {\small Universit\`a Cattolica del Sacro Cuore, Milano, Department of Mathematics, Finance and Econometrics} \\ {\small Email: gianpaolo.clemente@unicatt.it} \\
$^{\sharp}$  {\small University of Milano - Bicocca, Department of Statistics and Quantitative Methods} \\ {\small Email: rosanna.grassi@unimib.it}
}
\date{}

\maketitle

\begin{abstract}
In this article we propose a novel measure of systemic risk in
the context of financial networks. To this aim, we provide a
definition of systemic risk which is based on the structure, developed at different levels, of clustered neighbours around the nodes of the network. The proposed measure incorporates the generalized concept of clustering coefficient of order $l$ of a node $i$ introduced in \cite{CerCleGra}. Its properties are also explored in terms of systemic risk assessment. \\
Empirical experiments on the time-varying global banking network show the
effectiveness of the presented systemic risk measure and provide
insights on how systemic risk has changed over the last years, also
in the light of the recent financial crisis and the subsequent more stringent regulation for globally systemically important banks.
\end{abstract}


\section{Introduction}

The recent financial distress and its spread over the world economic
realities have pointed the attention of practitioners and academics
to the conceptualization and the management of systemic risk.
Indeed, even if the starting point of the crisis is well localized
in time and space -- 2008 in the US, with the failure of Lehman
Brothers -- the effects of this negative event have been and are
still pervasive quite everywhere.

The concept of systemic risk can be defined in a number of different
ways, also on the basis of the identification of the context under
investigation (see e.g. \cite{DeBandt}, \cite{Haldane}, \cite{Rochet}). Under a very general point
of view, systemic risk is the possibility that a negative occurrence
at a local level might generate a collapse at a global level. The
premise for the introduction of systemic risk is the definition of
a system, which is nothing but a unified structure composed by distinct
interconnected entities.

One of the most intuitive ways for modelling a systemic risk
framework is through complex networks (for a recent survey about networks
and systemic risk, we can refer to \cite{Caccioli2018} or \cite{Neveu2018}). In fact, a network is a
system composed by units -- the so-called nodes -- along with their
interconnections -- arcs or edges. Thus, systemic risk is the
possibility that an exogenous shock in one of the nodes triggers the
collapse of the entire network.

Therefore, it is not properly unexpected that several studies deal
with systemic risk problems in the framework of complex networks
(see e.g. \cite{Battiston2018}, \cite{Cont}, \cite{DiGangi},  \cite{Helfgott}, \cite{HubWal}
\cite{Torri}, \cite{Zhu}). 

The basis of systemic risk lies in the way in which shocks propagate
among the nodes of the network. Such a propagation is clearly
strongly dependent on the position and the density of the edges,
i.e. on the topological structure of the graph associated to the
network. Indeed, as intuition suggests, the presence of a large
number of interconnections leads to a more probable diffusion of the
local shocks, hence yielding a high level of systemic risk. In this
respect, it is worth mentioning \cite{DelliGatti2012}, \cite{billio12} and \cite{Markose}.

In the context of the relationship between interconnectedness and
systemic risk, a relevant role is played by the concept of
community. With the term community we refer to a set of nodes whose
mutual links are of particular strength (\cite{Fortunato2010}, \cite{Girvan2002}). Thus, the assessment of the community structure of a network
let understand how powerful the mutual interconnections among the
nodes are, hence providing useful insights on the systemic risk. This
argument suggests that the measurement of the entity of the
communities strength might represent a crucial step for the
exploration of systemic risk. In this respect, the clustering
coefficient of a network is of peculiar relevance.

The clustering coefficient of a given node is a relative measure of
the triangles including the considered node as a vertex with respect
to the hypothetical ones. Triangles are the easiest geometric
visualizations of the communities, since they offer the image of a
nonexclusive interaction among different agents.
Such a measure has been developed in all the cases of weighted,
unweighted, directed and undirected networks (see e.g. \cite{Barrat_2004}, \cite{CleGra}, \cite{Fagiolo_2007}, \cite{Onnela_2005}, \cite{WasFaust} and \cite{Watts_1998}). The extension of this community measure to the
overall network is obtained by simply taking the average of all the
clustering coefficients of the nodes.

In line with \cite{Bongini}, \cite{Minoiu} and \cite{Tabak}, this paper deals with the systemic risk assessment through the analysis of the clustering coefficient of a network.\color{black}
We move from a way to formalize a concept of community more general and informative than 
the standard clustering coefficient. In fact, communities might be appreciated
not only at the level of adjacent nodes, but also by considering
nodes which are at the periphery of the network. In so doing, we are
adding a stratification of the communities at different levels, in
order to capture the presence of community effects at different
distances from the considered nodes.

To this aim, we employ the high order clustering coefficient,
recently introduced by \cite{CerCleGra}. Furthermore, we adapt it 
to the systemic risk context in order to assess how each node is embedded in the entire system. In this way, we are providing specific indices that captures how a network is clustered at different levels. Additionally, a global systemic risk index is defined with the purpose of catching both the community structure around the node and the level of mutual interconnections of nodes that are at a specific geodesic distance.  In other words, it indicates the presence of high (or low) clustered areas, revealing parts of the network where the risk diffusion could spread easily. Furthermore, being this index defined as a weighted average of high order clustering coefficients measured at different levels, we are able to modulate the effects of both adjacent and peripheral nodes through the weights' distribution. Indeed, we can take into major consideration either the interactions of a node with its neighbours or the community structures generated by nodes at a wide distance from the considered one. 

Our theoretical proposal is validated over the paradigmatic example
of Global Interbank Network, which is particularly suitable for
our purpose. Indeed, the network structure of national interbank markets has been studied, at the global level, using the Bank of International Settlements (BIS) data set. Systemic risk is mainly related to the
interbank context and the considered empirical data allow to provide meaningful insights also to this type of literature (see, for instance, \cite{Bongini}, \cite{Garratt2011}, \cite{Giudici2016}, \cite{Giudici2017}, \cite{McGuire2006}, \cite{Minoiu}). 

We study the time-varying behaviour of the community structure of the global banking network over the sample period that goes from the first quarter of 2005 to the end of 2017. 
Data naturally induce a core-periphery network and then we focus on the behaviour of the core countries, i.e. countries whose banking systems report data to the BIS. In particular, to disentangle the role of core countries that host global systemically important banks (GSIBs), as defined in the list provided by the Financial Stability Board since november 2001, we will separately analyse the behaviour of countries in which at least a GSIB is present. In other words, we start by measuring the community structures at different levels on the basis of full global banking network's topology over each quarter of the year. Then, we compute high order clustering coefficients at a global level by focusing on two different subsets: the set of countries in which at least a GSIB is present and the set containing other core countries. In this way, we provide two alternative systemic risk indices that allow to assess the state of risk and to describe the pattern over time for both groups. \\
We observe that, in the entire period, countries, where a GSIB is present, show a clustering coefficient higher than other core countries,
confirming the important systemic role of the banks headquartered in these countries. Our analysis confirms evidence of a reduction of global banking connectedness as an effect of the cutback in cross-border lending, triggered by the subprime crisis and the subsequent sovereign debt crisis in the Euro area. Furthermore, results suggest a different pattern since 2011 between the two clusters. 
On one hand, other core countries show a tendency to diversify relationships as the average number of transactions increases and the average volume remains quite stable. On the other hand, lending countries, where a GSIB is present, are cutting their exposures in terms of numbers and volumes.
We interpret it as an effect of the \lq\lq systemic risk score'' introduced by the Basel Committee on Banking Supervision at the end of 2011 and the effectiveness of GSIBs regulation in inducing these banks to contain their systemic nature. \\
Furthermore, by providing a separate analysis between coefficients that considers only either in-flows or out-flows, we describe the different behaviour of countries in terms of risk-driver or risk-taker. It is worth mentioning how other core countries are more affected as risk taker. As regard to the \lq\lq out\rq\rq coefficient we observe a greater value for countries where a GSIB is present in the period 2009-2011. It is also noticeable the reduction of GSIB systemic impact over the last years. 

To sum up, the novelties of the present study are the following:
first, we construct a new systemic risk measure by using the concept
of stratified communities. In so doing, we include in the systemic
risk assessment also core-periphery effects in the analysis of
shocks propagation; second, we allow the tuning of the systemic risk
measure on specific levels of the communities, in order to give
credit to the action of peculiar parts of the considered network in
assessing systemic risk. In this, we are basically introducing a
family of systemic risk measures with a wide set of meanings and
interpretations; third, we provide a deep analysis of the Global
Interbanking System and infer aspects of the related systemic risk
which seem to remain unexplored in the standard frameworks. 

The rest of the paper is structured as follows. Preliminaries and notations are reported in Section \ref{Prel}. We summarize general notation in Subsection \ref{gennot}, while in Subsection \ref{hoclust} it is described the definition of the high order clustering coefficient introduced in \cite{CerCleGra}, which is the basis of the systemic risk measure defined below. In Section \ref{srm}, we provide a new indicator of systemic risk based on an extended version of the clustering coefficient. Some remarks on the systemic risk measure are in Subsection \ref{remark}.
By means of a small example, Section \ref{sim} stresses the potential of our proposal in respect to the classical weighted clustering coefficient defined in the literature. In Section \ref{na}, we provide a deep analysis of the interbank system. Conclusions follow.

\section{Preliminaries and notation}\label{Prel}

We here present the mathematical definitions supporting the structure of the paper, for the convenience of the reader.

\subsection{General notation}\label{gennot}
We denote by $G=(V,E)$ a graph, being $V$ the set of $n$ vertices
and $E$ the set of $m$ arcs (or edges), which are unordered pairs of
vertices. Vertices $i$ and $j$ are said to be adjacent when
$(i,j)\in E$. The degree $d_{i}$ of $i$ is the number of the edges incident upon $i$. A path connecting
vertices $i$ and $j$ is a sequence of distinct vertices and edges
between $i$ and $j$. If there exists a path between $i$ and $j$,
then $i$ and $j$ are said to be connected. The graph $G$ is
connected if all pairs of vertices of $G$ are connected.

The distance $ d\left( i,j\right) $
is the length of any shortest path connecting $i$ and $j$. Such a
shortest path is said to be a geodesic between $i$ and $j$ 
. All the geodesics between $i$ and $j$ have, of
course, the same length $d\left( i,j\right)=l$. We define the set
$\mathcal{G}_{ij}(l)$ as the one collecting all the geodesics
connecting the vertices $i$ and $j$; the generic element of
$\mathcal{G}_{ij}(l)$ is $g(l)=g_{ij}(l)$.

By conventional agreement, we assume that $ d\left(
i,j\right)=\infty $ when $i$ and $j$ are not connected. The diameter
of $G$, denoted by $diam(G)$, is an integer given by the length of
any longest path of $G$, and can be properly defined once $G$ is a
connected graph.

For a connected graph, we define the set: $$N_{i}(l)=\{j \in V |
d(i,j)=l\},$$ with $l=1,\dots,diam(G)$, and use the notation
$|N_{i}(l)|=d_{i}(l)$ to represent its cardinality.

If any edge $(i,j) \in E$ is associated to a positive real number
$w_{ij}$, then both the edges and the graph are weighted. Once we set that
$w_{ij}=0$ if and only if $(i,j) \notin E$, then we can describe
completely the edges of the graph through the real $n$-square matrix
$\mathbf{W}$ with entries $w_{ij}$, which is the weighted adjacency
matrix.
In particular, if $w_{ij}=1$ for all edges $(i,j) \in E$, then $\mathbf{W}$ is simply
the adjacency matrix $\mathbf{A}$ and the graph is unweighted. We will collapse this case into the more general weighted one.

The strength of vertex $i$ is the sum of the weights of the arcs
incident upon $i$. We denote it by $s_i$. Clearly, $s_i=d_i$ in the
unweighted case.

A weighted network is a graph with its weighted adjacency matrix.

The weight of a geodesic $g(l)$ between $i$ and $j$ is given by the
sum of the weights of its edges, and will be denoted by
$w_{ij}(l,g)$ hereafter. From this concept, we introduce the $l$-th
order strength of the node $i$ as
$$s_{i}(l)=\sum_{j \in N_{i}(l)} w_{ij}(l),$$
with $w_{ij}(l)=\min\limits_{g(l) \in
\mathcal{G}_{ij}(l)}\{w_{ij}(l,g)\}.$


When a direction is assigned to the edges of a graph $G$, then we
obtain a directed graph $D=(V,E)$, and $G$ represents the underlying
graph of $D$. The directed edges of $D$ are said arcs.

A directed path from $i$ to $j$ is a path whose arcs have the same
direction, which is the one going out from $i$ and going in $j$. The
existence of a directed path from $i$ to $j$ implies that $j$ is
reachable from $i$. Such a directed path is said to be an out-path
of $i$. In an intuitive way, one can say that the geodesic distance
$\overrightarrow{d}(i,j)$ from $i$ to $j$ is the length of a
geodesic out-path (or out-geodesic) connecting $i$ and $j$, and it
is set to $\overrightarrow{d}(i,j)=\infty$ when such an out-geodesic
does not exist.

By reverting the argument above, one has that the out-path of $i$
can be defined as an in-path of $j$, and we denote by
$\overleftarrow{d}(i,j)$ the length of any geodesic in-path (or
in-geodesic), with the usual agreement that
$\overleftarrow{d}(i,j)=\infty$ when such an in-path does not exist.

The directed graph $D$ is said to be strongly connected when all the
pairs of two vertices are mutually reachable. $D$ is said to be weakly connected if the underlying graph $G$ is connected.

By replacing $d$ with $\overrightarrow{d}$ and $\overleftarrow{d}$,
one can rewrite the definitions of $N_{i}(l)$, $d_{i}(l)$,
$\mathcal{G}_{ij}(l)$, $g(l)={g}_{ij}(l)$, $w_{ij}$, $w_{ij}(l,g)$,
$s_{i}(l)$, $w_{ij}(l)$ by $\overrightarrow{N}_{i}(l)$,
$\overrightarrow{d}_{i}(l)$, $\overrightarrow{\mathcal{G}}_{ij}(l)$,
$\overrightarrow{g}(l)=\overrightarrow{g}_{ij}(l)$,
$\overrightarrow{w}_{ij}$, $\overrightarrow{w}_{ij}(l,g)$,
$\overrightarrow{s}_{i}(l)$, $\overrightarrow{w}_{ij}(l)$ and
$\overleftarrow{N}_{i}(l)$, $\overleftarrow{d}_{i}(l)$,
$\overleftarrow{\mathcal{G}}_{ij}(l)$,
$\overleftarrow{g}(l)=\overleftarrow{g}_{ij}(l)$,
$\overleftarrow{w}_{ij}$, $\overleftarrow{w}_{ij}(l,g)$,
$\overleftarrow{s}_{i}(l)$, $\overleftarrow{w}_{ij}(l)$,
respectively.

\subsection{The high order clustering coefficient}\label{hoclust}

We now report the definition of the high order clustering
coefficient introduced in \cite{CerCleGra},
which is the basis of the systemic risk measure defined below.

In case of a weighted, undirected and connected graph $G$, we initially define a matrix
$\mathbf{P}(l)=[p_{ij}(l)]_{i,j \in V}$ for $l=1, \dots, diam(G)$, whose entries are
\begin{equation}
p_{ij}(l)= \begin{cases} \frac{w_{ij}(l)}{s_{i}(l)} & \mbox{if }  j
\in N_{i}(l) \mbox{ and } N_{i}(l) \neq \emptyset, \\ 0 &
\mbox{otherwise}.
\end{cases}
\label{pW}
\end{equation}
where $\mathbf{c}=[c_i]_{i \in V}$ is the
vector whose element $c_{i}$ is the weighted local clustering
coefficient of the node $i$ (see \cite{Barrat_2004}). If $l=0$, we
define $\mathbf{P}(l)=\mathbf{I}$, where $\mathbf{I}$ is the
identity matrix.

The local clustering coefficient of order $l$ is
$\mathbf{c}(l)=[c_i(l)]_{i \in V}$, obtained as
\begin{equation}
\mathbf{c}(l)=\mathbf{P}(l) \mathbf{c}. \label{Wclustering}
\end{equation}

\noindent If the graph is directed, weighted and weakly connected, matrix
$\mathbf{P}(l)$ in (\ref{pW}) becomes $\mathbf{\bar{P}}(l)$ with
entries:
\begin{equation}
\bar{p}_{ij}(l)= \begin{cases} \frac{\bar{w}_{ij}(l)}{\bar{s}_{i}(l)} &
\mbox{if }  j \in \bar{N}_{i}(l) \mbox{ and } \bar{N}_{i}(l) \neq \emptyset,\\
0 & \mbox{otherwise,}
\end{cases}
\label{eq:matrixPdirected}
\end{equation}
where:
\begin{itemize}
\item[$(a)$] $\bar{N}_{i}(l)=\overrightarrow{N}_{i}(l)$, $\bar{w}_{i,j}(l)=\overrightarrow{w}_{ij}(l)$ and $\bar{s}_{i}(l)=\overrightarrow{s}_{i}(l)$
in case only out-paths of node $i$ are taken into account, and
$\mathbf{\bar{P}}(l)=\overrightarrow{\mathbf{P}}(l)$;
\item[$(b)$]  $\bar{N}_{i}(l)=\overleftarrow{N}_{i}(l)$, $\bar{w}_{i,j}(l)=\overleftarrow{w}_{ij}(l)$ and $\bar{s}_{i}(l)=\overleftarrow{s}_{i}(l)$, considering only
in-paths of $i$, and
$\mathbf{\bar{P}}(l)=\overleftarrow{\mathbf{P}}(l)$;
\item[$(c)$]  $\bar{N}_{i}(l)=N_{i}(l)$, $\bar{w}_{i,j}(l)=w_{ij}(l)$ and $\bar{s}_{i}(l)=s_{i}(l)$ when all directions are considered, and
$\mathbf{\bar{P}}(l)=\mathbf{P}(l)$.
\end{itemize}
In all the previous cases, we assume $\mathbf{\bar{P}}(0)=\mathbf{I}$.

The local clustering coefficient of order $l$ in (\ref{Wclustering})
can be defined, respectively, for cases $(a)$, $(b)$ and $(c)$, as
follows:
\begin{equation}
\textbf{c}^{in}(l)=\overleftarrow{\mathbf{P}}(l) \textbf{c}^{in}.
\label{WDclusteringin}
\end{equation}
\begin{equation}
\textbf{c}^{out}(l)=\overrightarrow{\mathbf{P}}(l) \textbf{c}^{out}.
\label{WDclusteringout}
\end{equation}
\begin{equation}
\textbf{c}^{all}(l)=\mathbf{P}(l) \textbf{c}^{all}.
\label{WDclusteringall}
\end{equation}
where $\textbf{c}^{in}=[c^{in}_i]_{i \in V}$ and
$\textbf{c}^{out}=[c^{out}_i]_{i \in V}$ are vectors with entries
$c^{in}_{i}$ and $c^{out}_{i}$, respectively, that represent the in
and out weighted local clustering coefficients of node $i$, while
$\textbf{c}^{all}$ is the vector of local clustering coefficient for
graph $D$ (see \cite{CleGra}).

\section{Systemic risk measure}\label{srm}

We here propose a new indicator of systemic risk based on an extended
version of the clustering coefficient described in the previous
Section and which seems to be particularly effective for our
purpose.

If the case of an undirected graph, we introduce the vector
$\mathbf{h}=[h(l)]_{l=0, \cdots, diam(G)}$ such that
\begin{equation}
h(l)=\frac{1}{N}\sum_{i \in V} c_{i}(l). \label{h:clustering}
\end{equation}

Observe that $h(l)$ provides a feedback on how the nodes of the network are clustered together at a specific level $l$, being the mean of the clustering coefficients of order $l$. Then, the vector $\mathbf{h}$ collects the measures of all clusters. As a consequence, the distribution of its elements could give insights on the systemic risk of the network at each level $l$, as we will better be explained in the next Subsection. 

\noindent This suggests, as a quite natural further step, to use the elements of $\mathbf{h}$ to define a new measure of systemic risk. We define the index $h^\star$, \color{black} based on the
$h$'s, by taking their weighted mean as follows:
\begin{equation}
\label{hstar} h^\star=\sum_{l=0}^{diam(G)} x_lh(l),
\end{equation}
where $x_{l} \in [0,1]$ such that $\sum_{l=0}^{diam(G)} x_{l}=1.$

Notice that by means of $h^\star$ we are providing a specific systemic risk index that considers how each node is embedded in the network. In particular, the measure takes into account either the whole community structure around each node as well as the level of mutual interconnection of the nodes at a geodesic distance $l\geq 1$.  The presence of $x_{l}$ allows to introduce some flexibility in the computation of $h^\star$. Some remarks about possible distribution of $x_{l}$ are reported in Subsection \ref{remark}. Moreover, the effect of different weights $x_{l}$ will be tested in the numerical Section.\\
Also, observe that proposing this measure we are in line with the scientific debate on how clustering coefficient of a network might be viewed as a measure of systemic risk (see e.g. \cite{Minoiu} and \cite{Tabak}). 
 
Analogously to formula (\ref{h:clustering}), in the case of directed graph $D$, we can define
$\textbf{h}^{in}=[h^{in}(l)]_{l=0, \cdots, diam(G)}$,
$\textbf{h}^{out}=[h^{out}(l)]_{l=0, \cdots, diam(G)}$ and
$\textbf{h}^{all}=[h^{all}(l)]_{l=0, \cdots, diam(G)}$ where
\begin{equation}
h^{in}(l)=\frac{1}{N}\sum_{i \in V}   {c}_i^{in}(l), \qquad
h^{out}(l)=\frac{1}{N}\sum_{i \in V}   {c}_i^{out}(l), \qquad
h^{all}(l)=\frac{1}{N}\sum_{i \in V}   {c}_i^{all}(l),\label{hifrecce}
\end{equation}

and, in this case, the global clustering coefficients are:
\begin{equation}
h^{in,\star}=\sum_{l=0}^{diam(G)} x_l h^{in}(l), \qquad
h^{out,\star}=\sum_{l=0}^{diam(G)} x_l h^{out}(l), \qquad
h^{all,\star}=\sum_{l=0}^{diam(G)} x_l h^{all}(l). \label{hstarfrecce}
\end{equation} 


According to the literature which states that a high value of
clustering coefficient is associated to a high level of systemic
risk, we assume that the highest the value of $h^\star$'s in
(\ref{hstar}) and (\ref{hstarfrecce}), the highest the systemic risk
of the network.

\subsection{Some remarks on the systemic risk measure}\label{remark}

Notice that the coefficient $x_l$ in (\ref{hstar}) and
(\ref{hstarfrecce}) represents the ``weight'' to be assigned to $h(l)$ in the analysis of the entire
community structure related to the nodes of the network.As already pointed out, $h(l)$ brings information on the community structure at level $l$. 
In this respect, the selection of a specific distribution for the $x$'s
leads to different ways to include the periphery of the graph in the
systemic risk measurement. 

A high concentration at a specific level $l$ could indicate a particular attention to a more clustered area, revealing the intuition of the policy maker to focus on a part of the network where the risk diffusion could spread easily. For instance, if the mass of the weights $x$'s is
concentrated over the small values of $l$, then the considered systemic risk
measure will take into major consideration the community structures
close to the nodes of the graph. Differently, the case of
concentration of the $x$'s over large values of $l$ is associated to
a systemic risk more sensitive to communities far from the nodes.

The corner case $x_l=1$ reduces the vectors $\mathbf{h}$'s to the
clustering coefficients of order $l$. In this particular situation,
systemic risk measures $h^\star$'s take into consideration
communities at geodesic distance $l$ from the nodes. This analysis
might be relevant for the assessment of the stratified community
structure of the network. Indeed, a complete analysis of the corner cases $x_l=1$ with
$l=0,1,\dots, diam(G)$ allows to characterize the vulnerability of
the network and its ability of absorbing shocks by assessing the
presence of core-periphery communities. Notice that the very special case of $x_l=1$ for $l=0$ is the standard conceptualization of the clustering coefficient. \\

Moreover, the distinction between directed and undirected graphs
leads to remarkable differences in the definition of the systemic
risk measure of the network. When dealing with undirected graph,
strong communities are associated to high values of the weights
between the unordered couples of nodes forming an edge (see formula
(\ref{Wclustering})), and this property is clearly included in the
definition of the systemic risk measure, defined through formula (\ref{hstar}). In the directed case, weights should be intended
with a direction and the couples of nodes become ordered. Thus, a
node can be associated to a strong community in terms of in-paths
and a weak one when considering the out-paths (see
(\ref{WDclusteringin}) and (\ref{WDclusteringout})). Such a
characteristic of the nodes represents the basis of a concept of
``directed'' systemic risk measure (see $h^{in,\star}$ and
$h^{out,\star}$ in (\ref{hstarfrecce})), and this contributes to the
understanding of the vulnerability of the network. Indeed, a shock
occurring at a node $i \in V$ with a high level of $h^{out}_i$ is
expected to be rapidly propagated to the other nodes of the network,
while a high value of $h^{in}_i$ is associated to a probable
infection of node $i$ when the shock comes from outside it.


\subsection{Simulated example}\label{sim}

\begin{figure}[!h]
    \centering
    \includegraphics[scale=0.3]{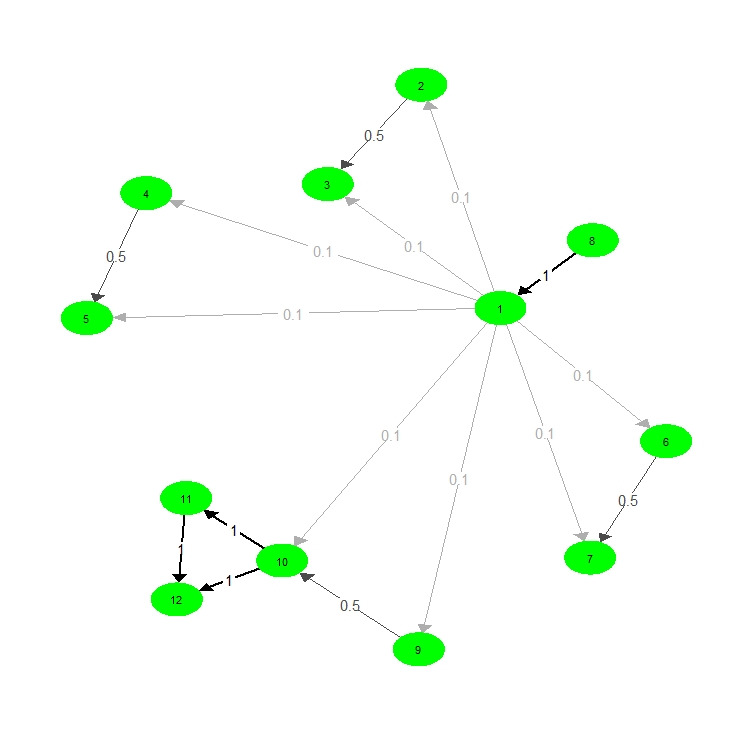}
    \caption{The simple weighted and directed graph $D$. Edges' opacity is proportional to weights.}
    \label{fig:Ex3}
\end{figure}

To show the effectiveness of the proposed high order clustering coefficients in capturing how a network is clustered at different levels -- hence leading to a a powerful definition of a systemic risk measure, as presented in the previous section -- we provide a simple weighted and directed graph $D$ of 12 nodes (see Figure \ref{fig:Ex3}). This allows to easily stress the potential of the index $h(l)$ in respect to the classical weighted clustering coefficient defined in the literature. 

We recall that the clustering coefficient should reflect the combined effect of the weights and the presence of triangles. Notice that, we focus on a directed graph since the vector of high order clustering
coefficients $\textbf{c}(l)$ of the underlying graph $G$ can be
obtained by simply rescaling the coefficients of $D$ (see \cite{CerCleGra} for details).


Since the graph is directed, we can refer to different kind of
geodesics (depending on whether the direction of the arcs is
considered or not).
We first compute the vector $\textbf{c}^{all}$ of coefficients as provided in \cite{CleGra}. We remind that each coefficient ${c}^{all}_{i}$ includes all kind of triangles which a node belongs to. \\
We then evaluate clustering coefficients of order
$l$ by means of formula (\ref{WDclusteringall}). Values are reported in
Table \ref{tab:CC3}. \\ 
 
\begin{table}[!h]
    \begin{tabular}{ c  c  c  c c c c  }
        \hline
        \hline
        Node & $\textbf{c}^{all}(0)$ & $\textbf{c}^{all}(1)$  & $\textbf{c}^{all}(2)$  & $\textbf{c}^{all}(3)$ \\

        \hline
        \hline

        1& 0.028 & 0.204 & 0.5 & 0  \\
        2 & 0.5 & 0.421 & 0.232 & 0.5  \\
         3 & 0.5 & 0.421 & 0.232 & 0.5  \\
         4 & 0.5 & 0.421 & 0.232 & 0.5  \\
         5 & 0.5 & 0.421 & 0.232 & 0.5  \\
         6 & 0.5 & 0.421 & 0.232 & 0.5  \\
         7 & 0.5 & 0.421 & 0.232 & 0.5  \\
        8 & 0 & 0.028 & 0.458 & 0.5 \\
        9 & 0.5 & 0.144 & 0.396 & 0  \\
        10 & 0.167 & 0.482 & 0.261 & 0 \\
        11 & 0.5 & 0.333 & 0.300 & 0.387 \\
        12 & 0.5 & 0.333 & 0.300 & 0.387 \\
        \hline
        $h^{all}(l)$ & 0.391 & 0.338 & 0.300 & 0.356  \\
        \hline
        \hline
    \end{tabular}
    \caption{Clustering coefficients and systemic risk indicators of order $l$ for the graph $D$ when all paths are considered (neglecting directions).}
    \label{tab:CC3}
\end{table}

\begin{table}[!h]
	\begin{tabular}{ c  c }
		\hline
		\hline
		$x_{l}$ & $h^{all,\star}$ \\		
		\hline
         Decreasing Weights & 	0.360 \\
         Uniform Weights & 0.346 \\
         Increasing Weights & 0.340 \\ 
		\hline
		\hline
	\end{tabular}
	\caption{Values of the global systemic risk measure $h^{all,\star}$ for different weights' distribution}
	\label{tab:CC4}
\end{table}

First, some remarks on the values of high order coefficients should be done. 
Looking at the clustering $\mathbf{c}^{all}(0)$, the node 1 has the
lowest positive value ($0.028$), being part of triangles with low
weights. This seems to suggest that node 1 is not remarkable in the
context of communities. However, this value does not reflect the
fact that the node 1 is adjacent to nodes with maximum clustering.
In other words, the interconnection with high clustered nodes is
softened by the presence of low weights. In addition, from node 1 we reach
nodes 11 and 12 in two steps, having clustering equal to 0.5. The
coefficients of order 1 and 2 ($c^{all}_1(1)=0.204$,
$c^{all}_1(2)=0.5$) are increasing, showing that an higher order of
clustering is able to capture the intensity of the communities
around the node at distances greater than 1.

Node 10 has the second lowest value ($c^{all}_{10}(0)=0.167$), but
the value of $c^{all}_{10}(1)$ is almost three times. This suggests
that the node forms communities with adjacent nodes with maximum
clustering.
It is worth noting the case of the node 8. This node does not contribute to form triangles, then its clustering coefficient is equal to zero. However, it has only one adjacent node and its direct connection with node 1 implies that the coefficient at level 1 completely absorbs the clustering value of the node 1. 
Also, clustering coefficients of order 2 and 3 are extremely high, due to
connections to high clustered nodes through geodesics of length 2
and 3. These aspects allow to interpret the position of this node in a completely different perspective, especially in spreading risk.

 The value of $h^{all,*}$ synthesizes the overall community structure of the network, thus providing a measure of the systemic risk associated to it. The choice of weights $x_{l}$ can modulate the intensity of the measure $h(l)$ in formula (\ref{hstarfrecce}), giving to this global network indicator a high degree of flexibility.
 
Here three possible scenarios for the weights $x$'s are considered:
\begin{itemize}
\item Decreasing weights $x_{l}=\frac{(l+1)^{-1}}{\sum_{l=0}^{diam(G)}(l+1)^{-1}}=\frac{(l+1)^{-1}}{H_{G}}$ where $H_{G}$ is the harmonic number of order $diam(G)+1$.
\item Uniform weights $x_{l}=\frac{1}{diam(G)+1}$
\item Increasing weights $x_{l}=\frac{(l+1)}{\sum_{l=0}^{diam(G)}(l+1)}=\frac{2(l+1)}{\left(diam(G)+1\right)\left(diam(G)+2\right)}$
\end{itemize}
For instance, assuming that weights $x_{l}$ are decreasing, we are reducing the impact of $h(l)$ respect the whole system when the distance $l$ increases. 

In this example, values of $h(l)$ state, on average, 
the presence of a strong community structure around the single node, as well as mutual interconnections at the maximal geodesic distance. As a consequence, both the closest 
and peripheral nodes have, on average, a similar influence on the node. In this regard, the distribution of weights is not very informative in this case. Indeed, different weights lead to very close values of $h^{all,\star}$ (see Table \ref{tab:CC4}).


Moving to the analysis of the directed case, 
\color{black}
we can separately investigate patterns of in or out-clustering by
using formulae (\ref{WDclusteringin}) and (\ref{WDclusteringout}).

Values are reported in Tables  \ref{tab:CC5} and   \ref{tab:CC4},
respectively. According to in-clustering, the vector
$\textbf{c}^{in}(0)$ considers in-triangles which a node belongs to.
Referring to the $1^{st}$ order, only two nodes ($11$ and $12$) are reachable
from a node with a positive in-clustering coefficient (i.e. the node $10$).
Notice that $c^{in}_{12}(1)$ is one half of $c^{in}_{11}(1)$ because
the node is also reachable from $11$.

Respect to the out-clustering, the node 8 is of interest; indeed, although its clustering coefficient is equal to zero, this node has positive coefficients of order $1$ and $2$, due to its connections via out-paths to nodes involved in out-triangles.

From the evaluation of the in- and out-patterns, a different behaviour in terms of systemic risk emerges, reflected by measures $h^{in,\star}$ and $h^{out,\star}$. 
The graph has a structure that seems more sensitive in receiving than spreading risk, and this characteristic is persistent also with different weights' distributions (Table \ref{tab:CC6}).
In particular, the community structure at level 0 favours the receiving of the risk (captured by the highest $h^{in}(0)$). There is the presence of few nodes that spread risk also to peripheral nodes.

\begin{table}[!h]
    \small
    \centering
    \begin{tabular}{ c  c  c  c c c c c c }
        \hline
        \hline
        Node & $\textbf{c}^{in}(0)$ & $\textbf{c}^{in}(1)$  & $\textbf{c}^{in}(2)$ & $\textbf{c}^{in}(3)$    & $\textbf{c}^{out}(0)$ & $\textbf{c}^{out}(1)$  & $\textbf{c}^{out}(2)$  & $\textbf{c}^{out}(3)$   \\

        \hline
        \hline

        1 & 0 & 0 & 0 &0  & 0.071 & 0.063 & 0 & 0  \\
        2 & 0 & 0 & 0 & 0 &  0 & 0 & 0 & 0 \\
        3 & 0.50 & 0 & 0 &0 & 0 & 0 & 0 & 0 \\
        4 & 0 & 0 & 0 & 0 &  0 & 0 & 0 & 0 \\
        5 & 0.50 & 0 & 0 &0 & 0 & 0 & 0 & 0 \\
        6 & 0 & 0 & 0 & 0 &  0 & 0 & 0 & 0 \\
        7 & 0.50 & 0 & 0 &0 & 0 & 0 & 0 & 0 \\
        8 & 0 & 0 & 0 &0 &0 & 0.071 & 0.063 &0\\
        9 & 0 & 0 & 0 & 0 &0 & 0.500 & 0 & 0\\
        10 & 0.50 & 0 & 0 &0  & 0.500 & 0 & 0 & 0\\
        11 & 0 & 0.50 & 0 &0 & 0 & 0 & 0 & 0 \\
        12 & 0.50 & 0.25 & 0 &0 &0 & 0 & 0 & 0 \\
        \hline
       Average & 0.208 & 0.063 &0 & 0 &0.048 & 0.053 & 0.005& 0 \\
        \hline
        \hline
    \end{tabular}
    \caption{Clustering coefficients and systemic risk indicators of order $l$ for the graph $D$ considering either in-paths or out-paths.}
    \label{tab:CC5}
\end{table}

\begin{table}[!h]
	\begin{tabular}{ c  c c}
		\hline
		\hline
		$x_{l}$ & $h^{in,\star}$ & $h^{out,\star}$\\		
		\hline
		Decreasing Weights & 	0.115 & 0.036 \\
		Uniform Weights & 0.068 & 0.026\\
		Increasing Weights & 0.035 & 0.015\\ 
		\hline
		\hline
	\end{tabular}
	\caption{Values of the global systemic risk measure $h^{in,\star}$ and $h^{all,\star}$ for different weights' distribution}
	\label{tab:CC6}
\end{table}

\section{Numerical Analysis}\label{na}
As in  \cite{Bongini}, \cite{Giudici2016} and \cite{Minoiu} we
designed a global banking network using the Bank for International
Settlements (BIS) consolidated statistics, which measure bank
exposures to different countries. These statistics capture
worldwide-consolidated claims of internationally active banks
headquartered in BIS reporting countries. In particular, we consider
international claims by a reporting country toward banks in
counterparty countries. In this way, we focus on the lending
activity of international banks. Here, nodes are countries and
weighted arcs represent positive cross-border exposures.

\begin{figure}[!h]
    \centering
    \includegraphics[scale=0.4]{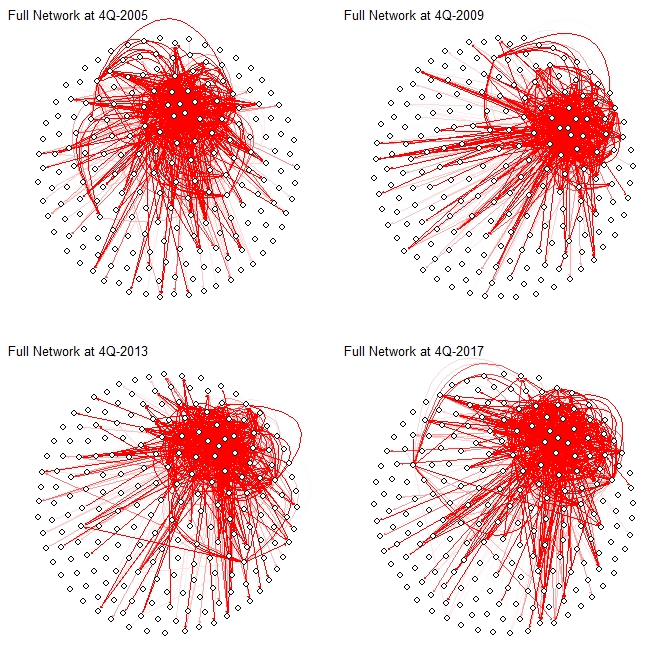}
    \caption{Cross-border global banking networks as on the end of 2005, 2009, 2013 and 2017. Red arrows represent arcs. Arcs opacity is proportional to weights (i.e., intensity of the exposures).}
    \label{fig:Net}
\end{figure}

We model each quarter of the year over the sample period (from the
first quarter of 2005 to the end of 2017) through a single network,
each one referred to links between banks of approximately 200
countries\footnote{The number of countries varies according to
different time-periods. Indeed, few isolated nodes are present at
specific times.}. \noindent Figure \ref{fig:Net} depicts the network
at four different time periods. Density is equal to 0.046 at the
$4^{th}$ quarter of 2005, then the network is sparse. It is
noticeable that networks become slowly denser over time (in terms of
the number of transactions): density is indeed equal to 0.055 at the
end of the time-period (4Q-2017). The number of arcs moves from 1540 to 2513 over
the sample period, while the number of countries remains stable. The
majority of countries is separated by at most two steps. Only very
few countries are reachable with three steps.

As also shown in Figure \ref{fig:Net}, data are designed so that the
resulting networks are characterized by core - periphery
structures\footnote{In network theory, a core-periphery structure
identifies a well-designed network model such that some nodes are
densely connected, whereas others are sparsely connected, in a
peripheral position (see \cite{Borgatti2000}).}. The core group
consists of countries whose banking systems report data to the BIS
in the analysed time-period, whereas other countries belong to the
periphery group. According to the BIS data, periphery countries are
analysed only as borrowers because for their banking systems only
information on inflows is available. Thus, the selection of the core
is strictly dependent on the data structure provided by the BIS.
Table \ref{tab:ListCountries} shows the list of 24 core countries
which reported to the BIS data about incoming and outgoing exposures
of international financial claims.

A specific analysis will regard also a subset of the core countries
that host global systemically important banks (GSIBs)\footnote{See
the FSB website for the updated list of G-SIBS banks
(http://www.fsb.org/2017/11/fsb-publishes-2017-g-sib-list/) and the
BCBS website for more information on how to assess systemically
important banks (https://www.bis.org/bcbs/gsib/)}. Since November
2011, the Financial Stability Board (FSB) releases the list of
G-SIBs each year, based on the Basel Committee on Banking
Supervision (BCBS) score of systemic risk (see \cite{BCBS2017} and \cite{BCBS2018}).  These banks are asked to
hold more capital (on top of Basel 3) and are subject to regulations
that are more stringent. In particular, we will define GSI
Countries, those countries in which at least a G-SIB is present (see
Table \ref{tab:ListCountries} for a list of GSIB countries).

\begin{table}[!h]
    \footnotesize
    \begin{tabular}{| l || c | c |}
        \hline
        Australia (AU)& Core & \\
        Austria (AT)&   Core     & \\
        Belgium (BE)&   Core    &GSI\\
        Canada (CA) &Core    & \\
        Chile (CL)  &Core    & \\
        China (CN)& & GSI\\
        China Taipei (TW) & Core & \\
        Finland (FI)&   Core     & \\
        France (FR) &Core   &GSI\\
        Germany (DE)&   Core&   GSI\\
        Greece (GR) &Core    & \\
        India (IN)  &Core    & \\
        Ireland (IE)&   Core     & \\
        Italy (IT)  &Core   &GSI\\
        Japan (JP)  &Core   &GSI\\
        Netherlands (NL)&   Core    &GSI\\
        Norway (NO) &Core    & \\
        Portugal (PT)   &Core    & \\
        Singapore (SG)  &Core    & \\
        Spain (ES)  &Core   &GSI\\
        Sweden (SE) &Core   &GSI\\
        Switzerland (CH)&   Core    &GSI\\
        Turkey (TR) &Core   & \\
        United Kingdom (GB) &Core   &GSI\\
        United States (US)  &Core   &GSI\\\hline
    \end{tabular}
    \caption{List of core countries and countries where at least one GSIB has its headquarters.}
    \label{tab:ListCountries}
\end{table}

\begin{figure}[!h]
    \centering
    \includegraphics[scale=0.3]{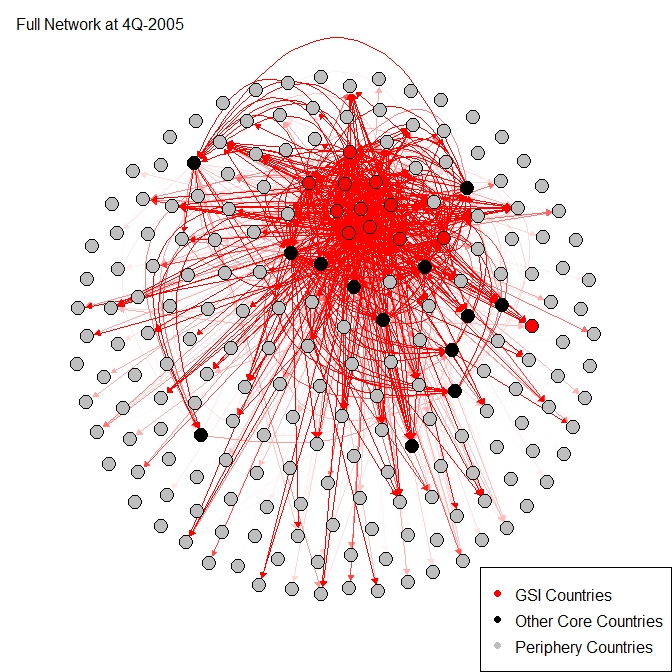}
    \includegraphics[scale=0.3]{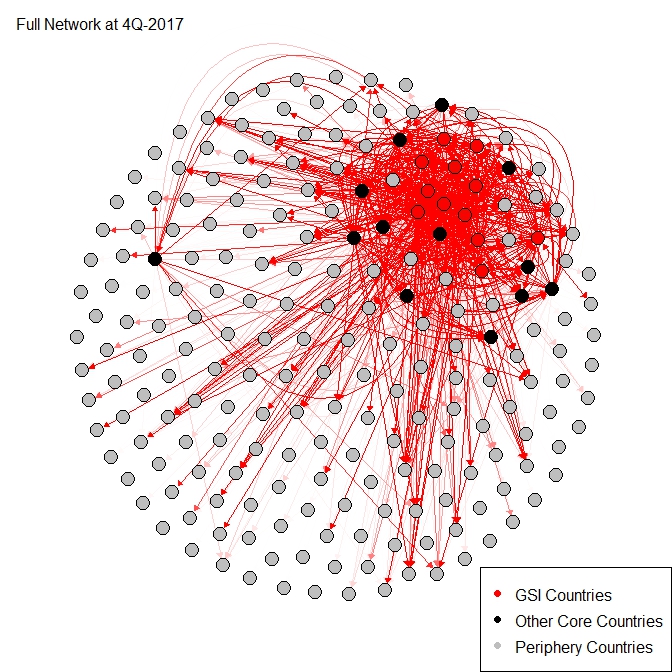}
    \caption{Cross-border global banking networks as on the end of 2005 and 2017 with the evidence of GSI, other core and periphery countries.}
    \label{fig:NetGS}
\end{figure}

In order to understand the role of G-SIB, we will separate countries in
three different clusters. In particular, we focus on GSI countries
and other core countries  (i.e. core countries that are not
classified as GSI). A third residual group considers periphery
countries. Figure \ref{fig:NetGS} gives an idea of the role of
different countries in the network at two different time periods. It
is remarkable that specific core countries become more integrated in
the dense part of the network by increasing, in particular, their
out-flows.

According to this classification, we are able to compute global
indicators based on a specific subset of countries. Initially, we
compute classical local clustering coefficients $c_{i}^{all}(0)$ for
the weighted and directed network (see \cite{CleGra}) and we
aggregate these local coefficients separately for GSI and other core
countries. Hence, we obtain two different estimates of $h^{all}(0)$ according to two different subsets of nodes.

\begin{figure}[!h]
    \centering
    \includegraphics[scale=0.35]{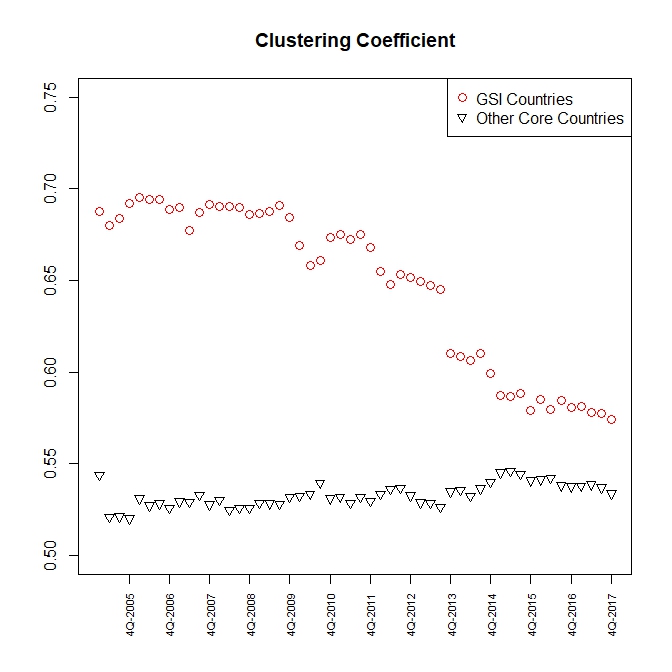}
    \caption{Global clustering Coefficients $h^{all}(0)$ computed by averaging local coefficients $c_{i}^{all}(0)$ at two different levels considering either only GSI countries or only other core countries respectively.}
    \label{fig:C}
\end{figure}

In the entire period, GSI countries show a clustering coefficient
higher than other core countries, confirming the important systemic
role of the banks headquartered in these countries (see Figure
\ref{fig:C}). Furthermore, results suggest a different pattern
between the two clusters of countries.  In particular, GSI countries
show a fall in clustering, starting from year-end 2008. This
reduction is in line with the general reduction of clustering for
the full network in 2009-2010, provided in \cite{Minoiu} and
ascribed to the perturbation in financial markets triggered by the
Lehman failure.

It is interesting to notice that financial communities tends to
weaken after the year 2011 for GSI countries. This phenomenon is
also more intense starting from 2013, the year when the BCBS revised
its methodology to assess GSIBs and the higher loss absorbency
requirement, to better comply with the purpose of reducing the
extent of failure of these banks.

\begin{figure}[!h]
    \centering
    \includegraphics[scale=0.4]{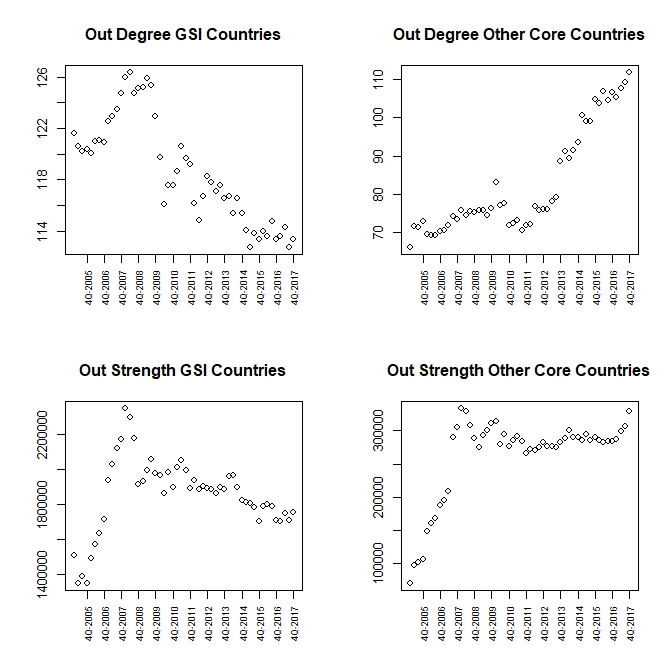}
    \caption{Out-degree and out-strength for GSI and other core countries.}
    \label{fig:Out}
\end{figure}

This behaviour is partially explained by the pattern of out-degree
and out-strength reported in Figure \ref{fig:Out}. There is a tendency in the network, particularly with other core
countries, to diversify relationships as the average number of
transactions increases and the average volume remains quite stable.
The same pattern is not observed for GSI countries. In particular,
considering the volume of transactions, evidence suggests that GSI
countries are cutting their exposures to almost all partners. This
behaviour provides an evidence of the effectiveness of the
regulation, dampening  the GSIB systemic impact.

\begin{figure}[!h]
    \centering
    \includegraphics[scale=0.35]{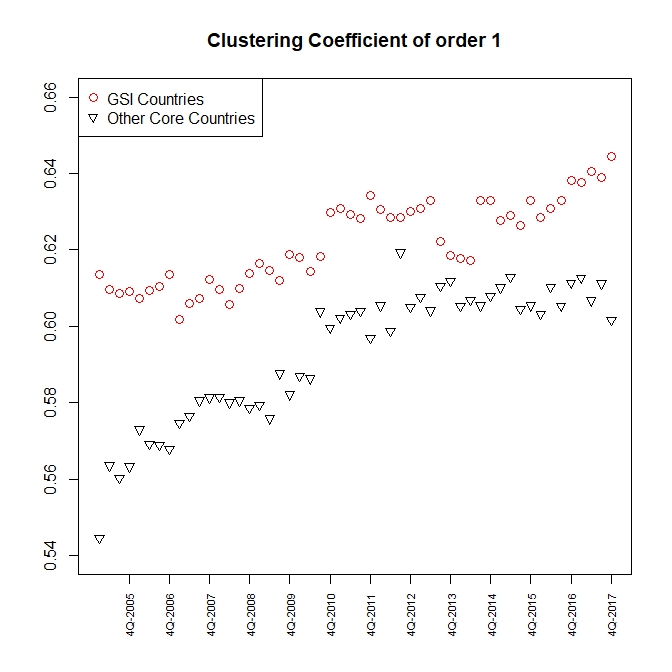}
    \caption{Global clustering coefficients of order 1 $h^{all}(1)$ computed by averaging local coefficients $c_{i}^{all}(1)$ at two different levels considering either only GSI countries or only other core countries respectively.}
    \label{fig:C1}
\end{figure}

In order to test the behaviour of clustering coefficients of order
$l$, we compute $\textbf{c}^{all}(1)$ by means of formula
(\ref{WDclusteringall}). In order to obtain two values of the
synthetic indicator, one for GSI subset and one for other core
countries subset, we average local coefficients of order $1$ of
countries belonging to the same subset. Results are reported in
Figure \ref{fig:C1}.

We examine, in this way, the systemic risk at different observation
scales. While classical clustering coefficient $c^{all}_{i}$ takes
into account how the node $i$ form communities in the context of the
overall system, $c^{all}_{i}(1)$ measures how much the neighbours of
node $i$ form triangles and communities. Patterns of Figure
\ref{fig:C1} show that GSI countries are also connected to
well-established communities confirming the central role of these
countries in term of systemic risk.

It is worth mentioning that GSI countries, against the reduction of average number and volume of transactions, maintain a high-level of communities, captured by $h^{all}(1)$, due to their relations with neighbours that belong to many triangles. 
This effect is partially induced by the contraction of lending activity towards periphery countries, leading to an increase of clustering
coefficient of order 1.

\begin{figure}[!h]
    \centering
    \includegraphics[scale=0.3]{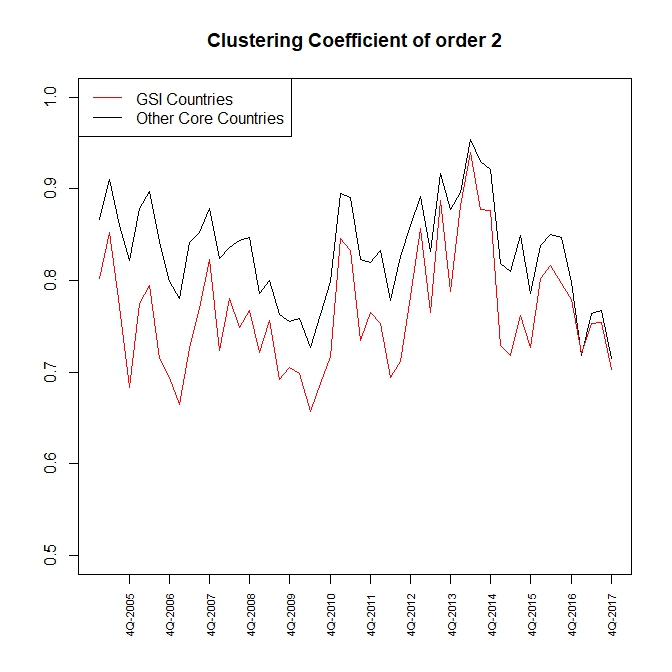}
        \includegraphics[scale=0.3]{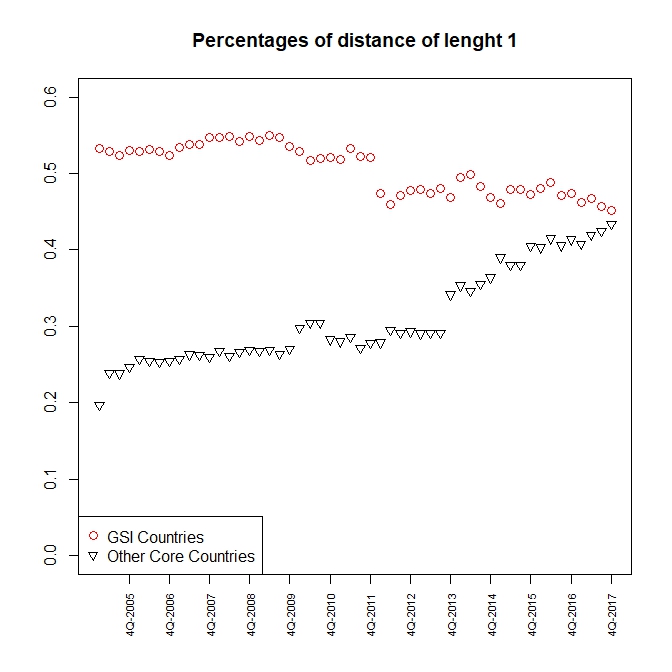}
    \caption{On the left side, Figure shows global clustering coefficients of order 2 $h^{all}(2)$ computed by averaging local coefficients $c_{i}^{all}(2)$ at two different levels considering either only GSI countries or only other core countries respectively. On the right side, Figure displays the percentage of geodesic of length 1 with respect to the total for each cluster (GSI and other core countries)}
    \label{fig:C2}
\end{figure}

On the left side, Figure \ref{fig:C2} depicts global clustering
coefficients of order 2 ($h^{all}(2)$) computed for the two subsets of countries. A
similar pattern between GSI and other core countries is observed, specially over the last period. This
behaviour can be partially explained by observing the ratio of
geodesics of length 1 to the total number of geodesics for each
cluster of countries (Figure \ref{fig:C2}, right side). Indeed, as
already seen in Figure \ref{fig:Out}, the average number of
borrowers for the two subsets of lenders behaves in a opposite
manner since 2008-2009. On one hand, GSI countries reduced the
average number of transactions, on the other hand, other core
countries increase it. At the end of the period, very close
out-degrees are observed for both subsets, so that the percentage of
nodes at distance 1 becomes very similar and clustering coefficient
of order 2 is obviously affected by this fact.
Additionally, the similarity of the 2-order clustering coefficients displays that both subsets are connected in two steps to comparable communities.

\begin{figure}[!h]
    \centering
    \includegraphics[scale=0.3]{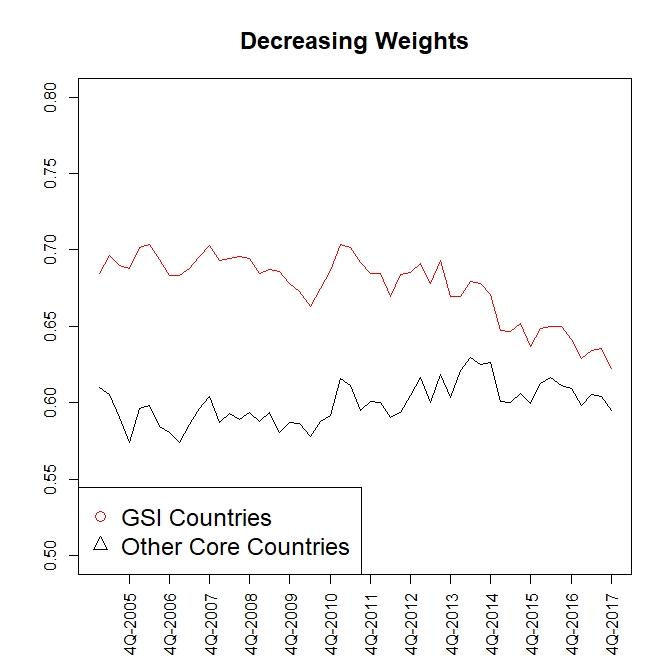}
     \includegraphics[scale=0.3]{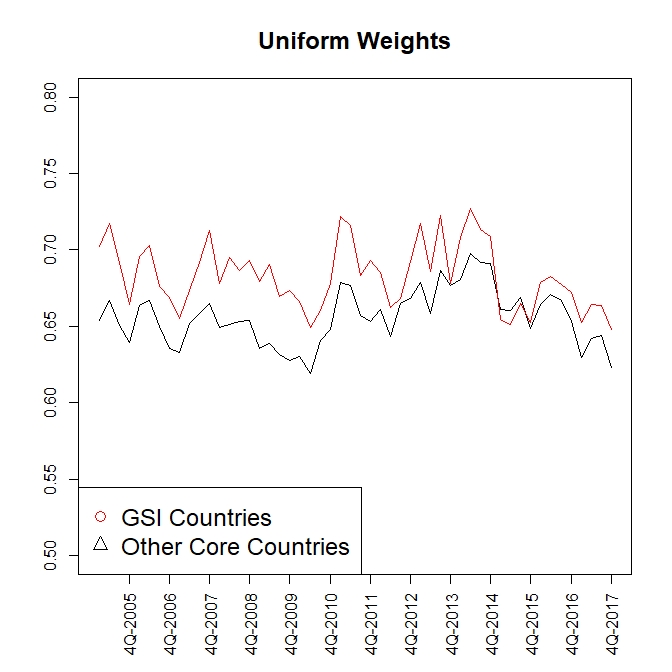}
     \includegraphics[scale=0.3]{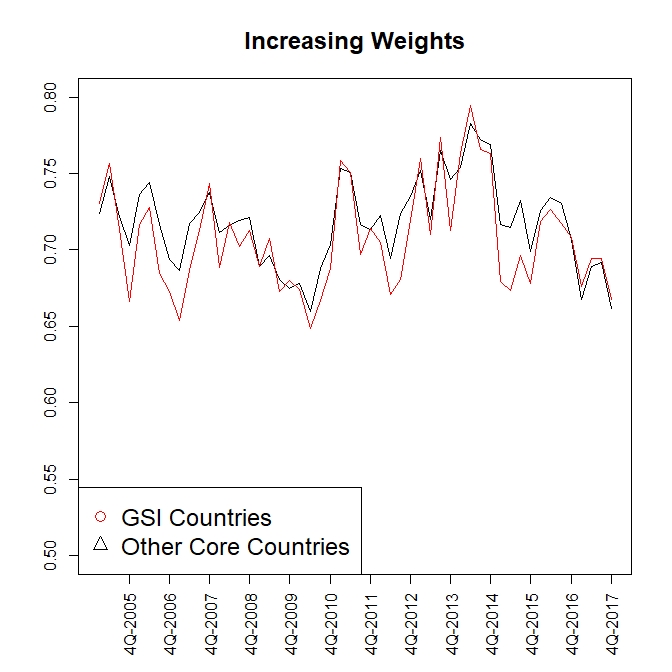}
    \caption{Figure reports the high order clustering coefficient for different choices of weights, where $\textbf{h}^{all,*}$ is computed by averaging local coefficients for the two subsets of countries.}
    \label{fig:HAverage}
\end{figure}

\newpage
Figure \ref{fig:HAverage} shows the high order clustering $\textbf{h}^{all,*}$ computed by considering different choices of weights,  for both subsets of countries. Specifically, we test here the same distributions of weights described in Section \ref{sim}, reporting the results in Figure \ref{fig:HAverage}. 
A different weights' concentration produces distinct patterns, in terms of comparison between GSI and other countries. 
Focusing, for instance, on a decreasing distribution, it is confirmed the prominent role of GSI countries in spreading and receiving risk to (and from) their neighbours. On the contrary, assigning more weight to the relationships between nodes at higher distance, other core countries tend to have a pattern in line with GSI countries. 
Hence, we provide here a different view of systemic risk: unless GSI countries surely play a key role in spreading and receiving risk, by looking beyond the adjacent nodes, we deduce that also banks of other core countries can significantly contribute to risk diffusion. Furthermore, independently from the weights' distribution, in all cases, we have that the behaviour of both subsets tends to be aligned over the last years.

\newpage

\begin{figure}[!h]
	\centering
	\includegraphics[scale=0.3]{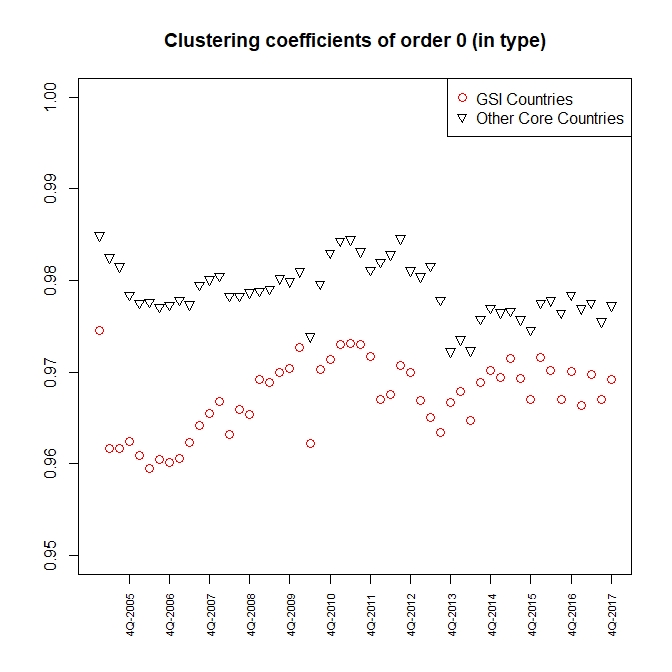}
	\includegraphics[scale=0.3]{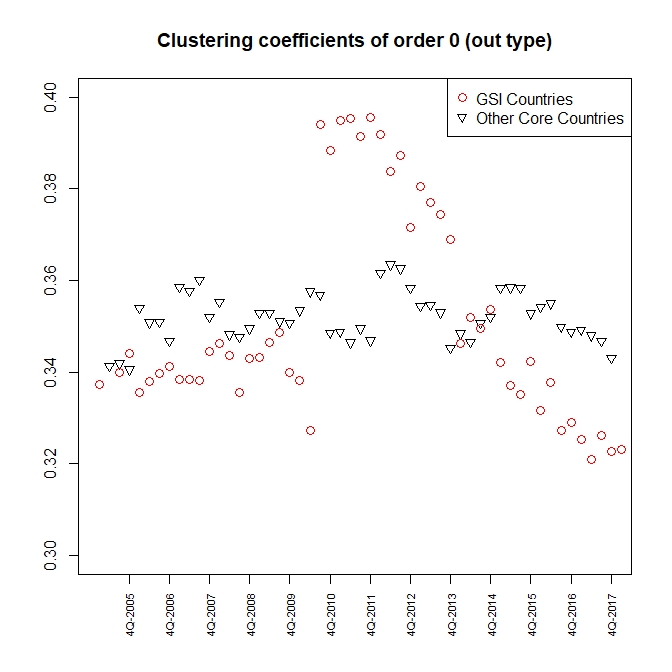}
	\caption{In and out clustering coefficients $h^{in}(0)$ and $h^{out}(0)$. Both coefficients are computed by averaging local coefficients considering either only GSI countries or only other core countries respectively.}
	\label{fig:InOutC}
\end{figure}

The numerical analysis has been also extended by considering in a separate way only either in-paths or out-paths. In this way, we catch only the effect of community structures in spreading or receiving risk. To this aim, we report in Figure \ref{fig:InOutC} values of $h^{in}(0)$ and $h^{out}(0)$ for both subsets of countries. It is worth mentioning the different pattern between other core and GSI countries. The former ones tend to have higher connections of the in-type to their neighbours. GSI countries have instead, on average, an higher role in spreading risk towards their adjacent nodes. In particular, it is noticeable the specific pattern $h^{out}(0)$ for GSI countries since the end of 2011. Indeed, the structure of financial communities of out-type tends to weaken, with a significant decrease from 2013, probably due to the reaction of banks in these countries to the systemic risk regulation. \\

\begin{figure}[!h]
	\centering
	\includegraphics[scale=0.3]{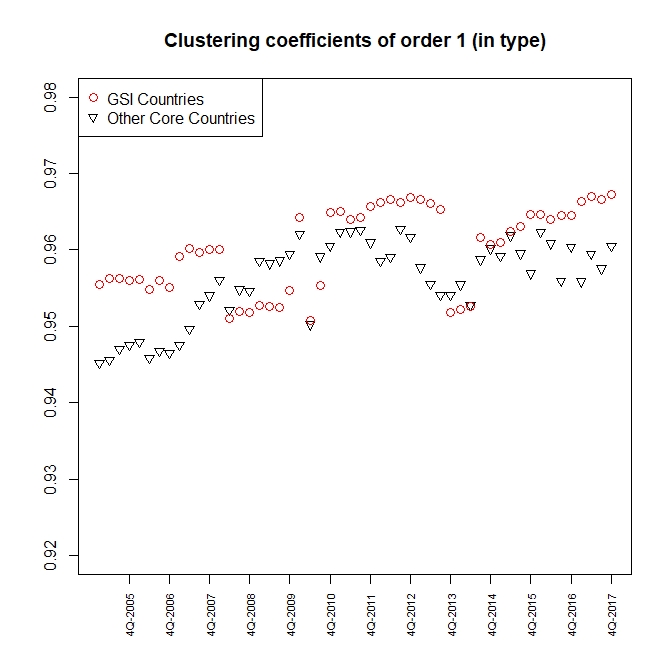}
	\includegraphics[scale=0.3]{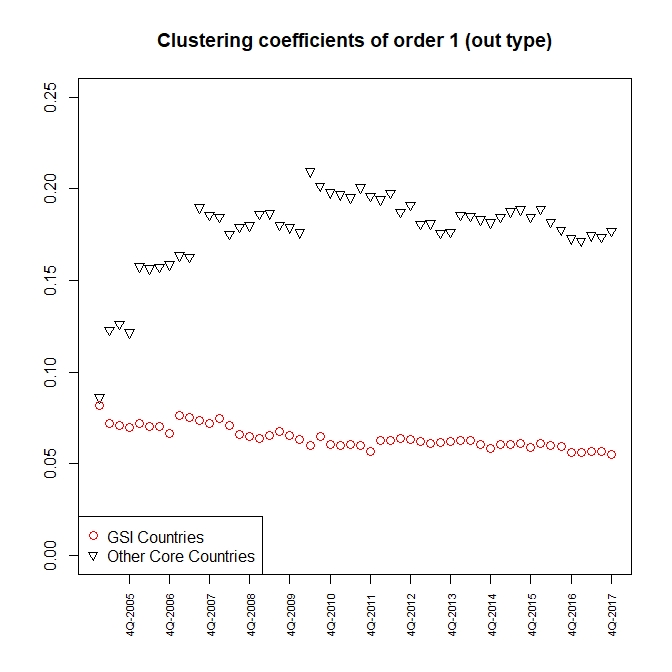}
	\caption{In and out clustering coefficients of order 1, $h^{in}(1)$ and $h^{out}(1)$. Both coefficients are computed by averaging local coefficients considering either only GSI countries or only other core countries respectively.}
	\label{fig:InOutC1}
\end{figure}

As regard to level 1, we have a very different picture of the network (see Figure \ref{fig:InOutC1}). GSI countries tend to show an higher level of connections of the in-type with strong community structures. On other hand, GSI countries spread, on average, risk toward countries that have lower out-clustering coefficients  $h^{out}(0)$. So it is interesting to note that, in this case, there is only a low further propagation of risk. On other hand, we observe an increasing pattern of $h^{out}(1)$ for core countries over time showing that banks of these countries are more and more connected to banks of risk-giver countries. 

\begin{figure}[!h]
    \centering
    \includegraphics[scale=0.3]{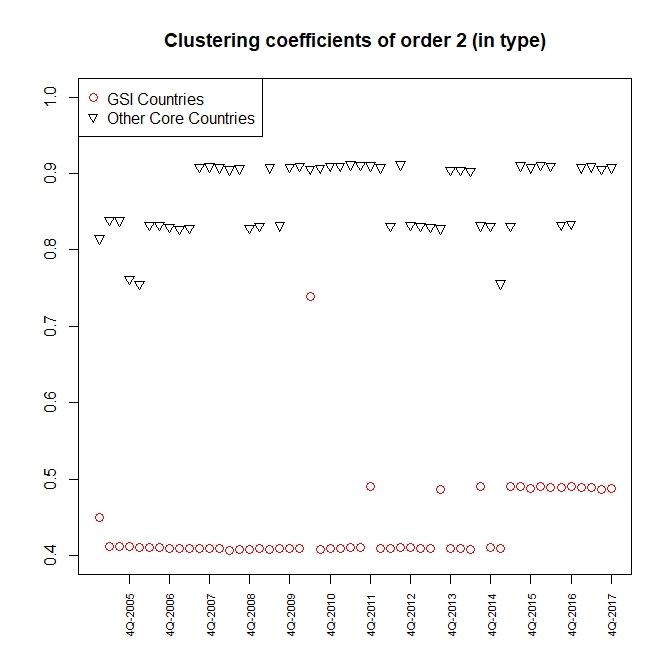}
    \includegraphics[scale=0.3]{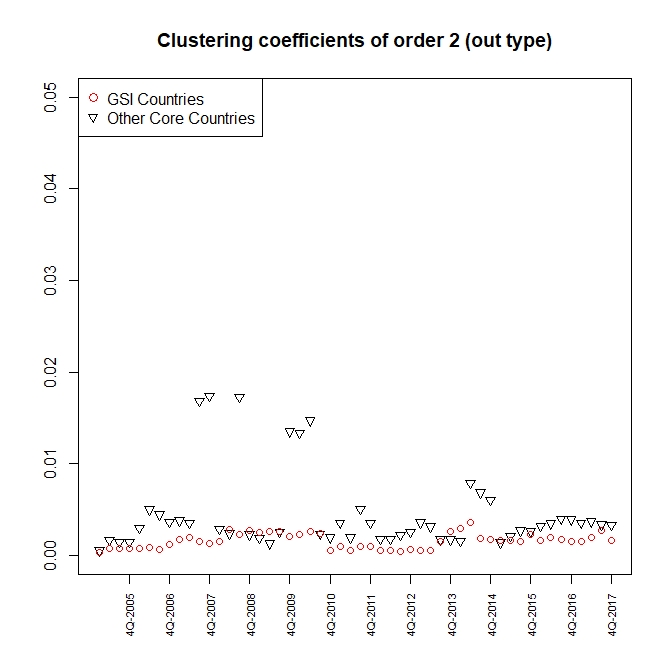}
    \caption{In and Out Clustering of order 2,  $h^{in}(2)$ and $h^{out}(2)$. Both coefficients are computed by averaging local coefficients considering either only GSI countries or only other core countries respectively.}
    \label{fig:InOutC2}
\end{figure}

\begin{figure}[!h]
    \centering
    \includegraphics[scale=0.3]{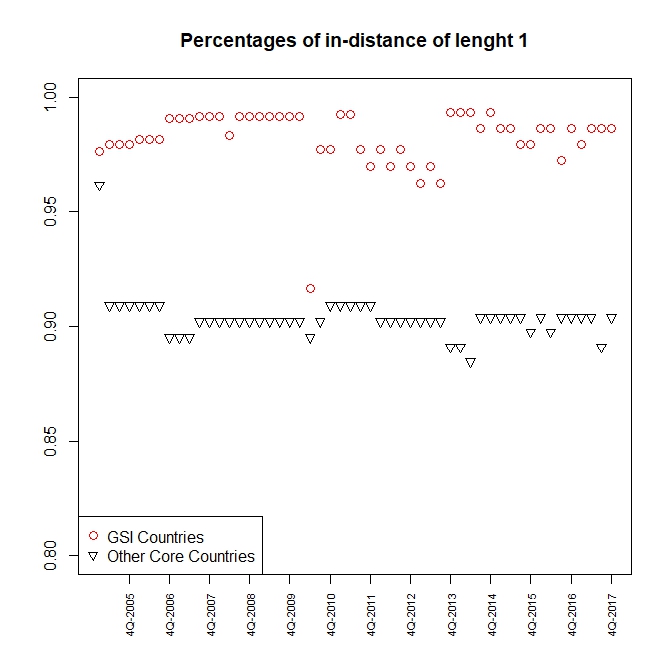}
    \includegraphics[scale=0.3]{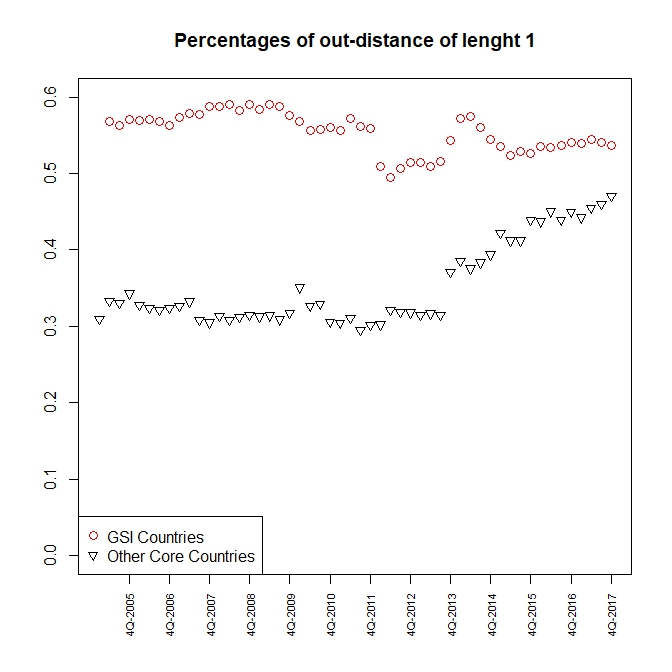}
    \caption{Percentages of directed (in and out respectively) geodesic of length 1 with respect to the total for each cluster (GSI and other core countries)}
    \label{fig:Dist}
\end{figure}

Values of $h^{in}(2)$ and $h^{out}(2)$ have been reported in Figure \ref{fig:InOutC2}. In these networks, the maximal length of in-geodesics is equal to 2. However, we have a very low proportion of countries (see Figure \ref{fig:Dist}, left side) that are reachable in two steps via an in-path\footnote{The ratio of in-geodesics of length 1 to the total number of in-geodesics is equal to 96\% for GSI countries and around 90\% for other core countries.}. So, differences observed for the $h^{in}(2)$ clustering between the two subsets are mainly motivated by the behaviour of specific countries. On average, other core countries, when act as borrowers, are more connected, via in-paths of length 2, to strong community structure.
Concerning the out-clustering of level 2, all core countries show very low connections. Although, over time, the percentages of out-distances of length 2 is reducing for GSI countries and increasing for other core countries, all countries are, on average, mostly connected at distance 2 with peripheral countries characterized by low risk.

\begin{figure}[!h]
	\centering
	\includegraphics[scale=0.3]{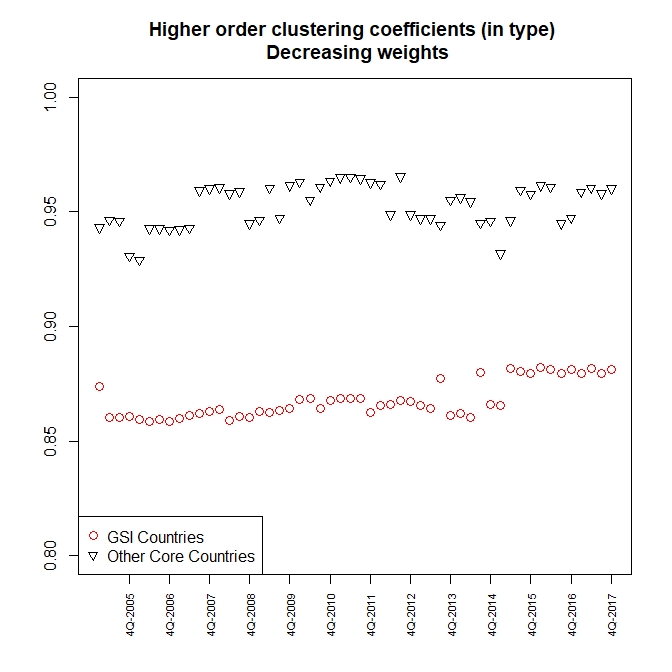}
	\includegraphics[scale=0.3]{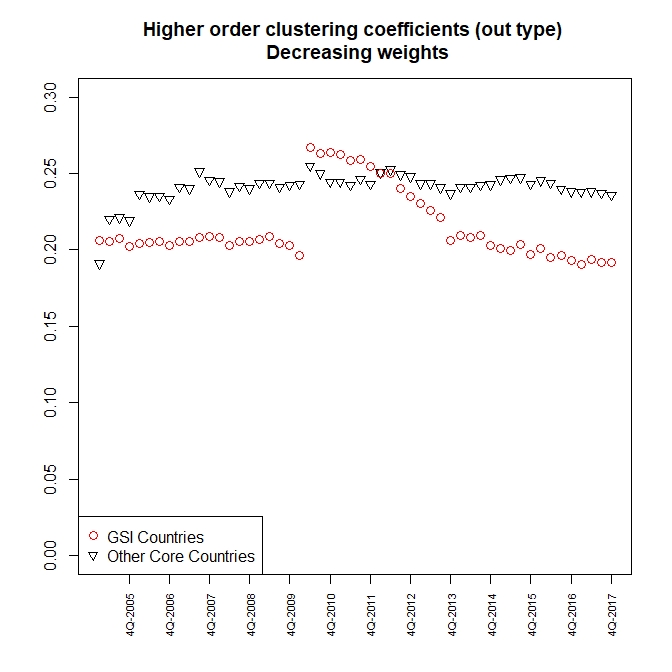}
	\includegraphics[scale=0.3]{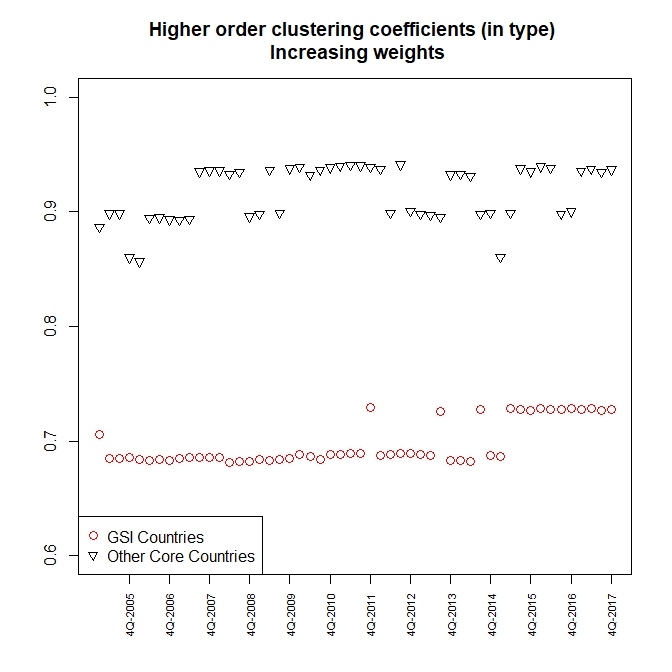}
		\includegraphics[scale=0.3]{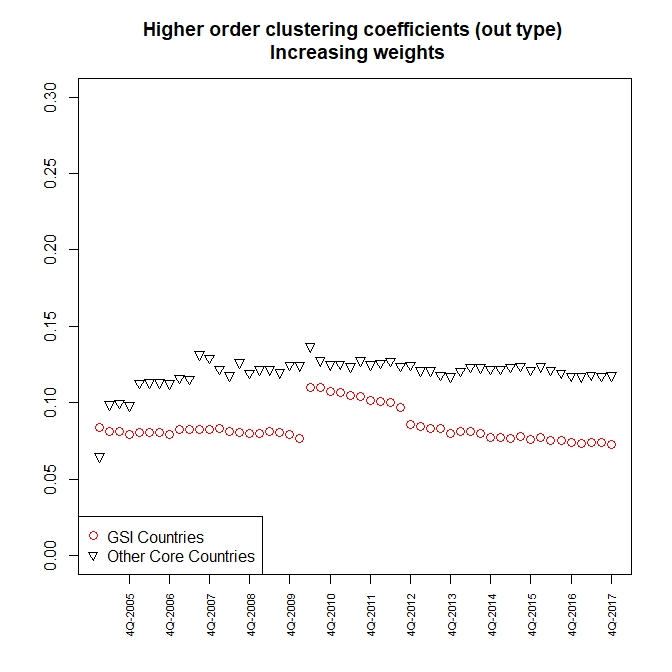}
	\caption{ In and out high order clustering coefficients $h^{\star,in}$ and $h^{\star,out}$. Coefficients are computed for both subsets of countries and with different weights.}
	\label{fig:InOuthstar}
\end{figure}

 Now, we focus on the systemic risk measures $h^{in,\star}$ and $h^{out,\star}$ regarding in and out-flows respectively (see Figure \ref{fig:InOuthstar}). As stressed in \cite{Tabak}, an higher clustering coefficient of the in-type may reflect higher systemic risk because failure of the borrowing node in an “in” triangle can trigger simultaneous non-repayments to the lending nodes, and this can make them unable to honour their own obligations. We show that both the in-clustering and the $h^{in,\star}$ assess the high state of stress in the network. It is worth mentioning how other core countries are more affected as risk taker than GSI countries. This effect is more evident when weights are more concentrated on adjacent nodes because of the high level of interaction of these countries as borrowers. It is instead more noticeable the effect of weights' distribution on GSI countries. We have indeed that values of $h^{in,\star}$ for this subset are decreasing approximatively from 0.87 to 0.7 when increasing weights are chosen. These countries are less affected, when they act as borrowers, by the effect of countries that are at higher distances. Differently focusing on $h^{out,\star}$, we observe a greater value for GSI countries in the period 2009-2011. It is also confirmed the reduction of GSIB systemic impact over the last years, mainly characterized by the reduction of the exposure of GSI countries.

\newpage
\section{Conclusions}

Systemic risk in finance is a concept not easy to be formalized through a quantitative measure and a huge and fast growing literature is interested in this issue.
A quite natural approach is based on the use of complex networks.  \\
In a financial system, the interconnectedness among entities plays a fundamental role in situations of distress. Moving from this fact, we exploit the concept of community, usually relevant in understanding the relationship between interconnectedness and systemic risk.
In particular, we consider a generalization of the concept of clustering coefficient in order to catch both the presence of clustered areas around a node and/or  high levels of mutual interconnections at different distances from the node itself. We provide a new systemic risk measure computed as the weighted average of high order clustering coefficients at different levels. On one hand, this proposal leads to a synthetic indicator able to assess the general state of stress of the financial system. On the other hand, the distribution of weights allows to introduce a degree of flexibility, in order to modulate the effects of both adjacent nodes and peripheral nodes. \\
An empirical application to time-varying global banking network is developed. Results show the effectiveness of these measures in reflecting how systemic risk has changed over the last years, also in the light of the recent financial crisis. Furthermore, we emphasize a different pattern of behaviour between countries where a GSIB is headquartered and other core countries, more noticeable since 2013. This effect, that could be interpreted as a reaction to the specific regulation inducing banks to contain their \lq\lq systemic\rq\rq nature, is in line with the recent report by the Committee on the Global Financial System \cite{CGFS2018} that shows that GSIBs become more selective and have also repositioned themselves toward less complex activities, as a response to the regulatory reforms process that is under way.

\bigskip

\bibliographystyle{plain}


\end{document}